\documentclass[journal]{IEEEtran}
\usepackage{amsmath}
\usepackage{lipsum}
\usepackage{graphicx}
\usepackage[justification=centering]{caption}
\usepackage{epstopdf}
\usepackage{tabularx}
\usepackage{algorithm}
\usepackage{algorithmic}
\usepackage{amssymb}
\usepackage{xcolor}

\usepackage[hyphens]{url} 
\usepackage{lipsum}

\ifCLASSINFOpdf
\else
\fi
\begin{document}
%
\title{Edge-enabled Metaverse: The Convergence of Metaverse and Mobile Edge Computing}

%

\author{Sahraoui Dhelim, Tahar Kechadi, Liming Chen, Nyothiri Aung, Huansheng Ning and Luigi Atzori.

	\thanks{Sahraoui Dhelim, Tahar Kechadi and Nyothiri Aung are with the School of School of Computer Science, University College Dublin, Ireland.} 
	\thanks{Liming Chen is with the School of Computing, Ulster University, U.K.}
	\thanks{Luigi Atzori is with the Department of Electrical and Electronic Engineering, University of Cagliari, Italy}
	\thanks{Huansheng Ning is with the School of Computer and Communication Engineering, University of Science and Technology Beijing, China.}
	\thanks{Corresponding author: Huansheng Ning (ninghuansheng@ustb.edu.cn).}
}

%
%

\markboth{Journal of \LaTeX\ Class Files,~Vol.~14, No.~8, August~2015}%
{Shell \MakeLowercase{\textit{et al.}}: Bare Demo of IEEEtran.cls for IEEE Journals}
%



\maketitle

\begin{abstract}
The Metaverse is a virtual environment where users are represented by avatars to navigate a virtual world, which has strong links with the physical one. State-of-the-art Metaverse architectures rely on a cloud-based approach for avatar physics emulation and graphics rendering computation. Such centralized design is unfavorable as it suffers from several drawbacks caused by the long latency required for cloud access, such as low quality visualization. To solve this issue, in this paper, we propose a Fog-Edge hybrid computing architecture for Metaverse applications that leverage an edge-enabled distributed computing paradigm, which makes use of edge devices computing power to fulfil the required computational cost for heavy tasks such as collision detection in virtual universe and computation of 3D physics in virtual simulation. The computational cost related to an entity in the Metaverse such as collision detection or physics emulation are performed at the end-device of the associated physical entity. To prove the effectiveness of the proposed architecture, we simulate a distributed social metaverse application. Simulation results shows that the proposed architecture can reduce the latency by 50\% when compared with the legacy cloud-based Metaverse applications.

\end{abstract}


%
\IEEEpeerreviewmaketitle

\section{Introduction}
%
%
%
%
\IEEEPARstart{T}{he} Metaverse has emerged as prominent candidate to replace the Internet as we know it today. Instead of navigating through webpages through 2D flat devices like smartphone or laptop, the Metaverse allows users to interact with 3D virtual worlds using their avatar where they can work, study or just have fun in a virtual universe. Thanks to virtual visualization technologies such as Virtual Reality (VR), Augmented Reality (AR), Mixed Reality (MR) and Extended Reality (XR) the users can enjoy immersive virtual experience enriched with realistic feedback sensors. The Metaverse have received notable attention, with many multi-billion companies willing to invest in its applications. Facebook, which recently changed its name to Meta, reveled their plan to invest \$10 billions in Metaverse development \cite{Kastrenakes2022}. Similarly, Microsoft spent \$70 billion in the acquisition of Metaverse company Activision Blizzard \cite{Kalogeropoulos2022}. 

The Metaverse is expected to revolutionize traditional Internet and dominate all aspects of our lives. It has promising applications in virtual education, virtual workplace, real estate, to name a few. However, to fulfill the potentials of massive virtual universe, the Metaverse applications must adhere to certain requirements such as ultra-low latency, massive resource demands, interoperability between applications, and security and privacy issues \cite{dionisio20133d}.

To date, Metaverse implementations use centralized cloud-based approach for avatar physics emulation and graphical rendering. Such centralized design is unfavorable as it suffers from several drawbacks caused by the long latency required for cloud access, such as low quality visualization. For example, Second Life's has many virtual worlds that depend on centralized architecture \cite{thompson2010next}, each virtual world is divided into smaller regions, each managed by dedicated server. Most of the computational required for running the virtual world simulation such as physics emulating, 3D animation and collision detection are performed by the region's centralized server. Therefore, the number of users that can access each region is limited by the server's computational and communication capacity. Such limitation defeats the purpose of universal virtual world, which is supposed to accommodate avatars at least as much as the real world location can physically accommodate people. Fog Computing \cite{Hu2017} and Mobile Edge Computing \cite{Abdenacer2021} has been proven effect to tackle the issues faced by cloud-based systems, by moving the computational heavy load near to the end-user and distribute it among edge devices; such approach can significantly reduce the latency and optimize the system performance.  Therefore, in this paper, we propose a Fog-Edge hybrid computing architecture for Metaverse applications that leverage an edge-enabled distributed computing paradigm, which makes use of edge devices to fulfil the required computational for heavy tasks such as collision detection in virtual universe and computation of 3D physics. The computational costs of each entity such as collision detection or physics emulation are performed at the end-device of that physical entity. For example, for every avatar in the Metaverse, the user's edge devices are responsible to perform computation tasks related to the avatar's physical movement such as its momentum, mass and physical forces of surrounding entities. Our contribution can be summarized as follow:

\begin{itemize}
    \item Discuss the development trend of Metaverse, and its similarity with Blockchain development history, and the possibility that we end-up with multi fragmented Metaverses rather than a distributed universal Metaverse. 
    \item Propose a hybrid Fog-Edge architecture for Metaverse applications that leverages the physical entities' end-devices computing resources to perform the computational tasks required to run the Metaverse.
    \item Simulate the proposed architecture in a distributed social Metaverse application use-case scenario to evaluate the improvements with respect to a centralized approach.
\end{itemize}

The rest of the paper is organized as follows:
Section \ref{sec.2} reviews recent research literature of Metaverse technologies. In Section \ref{sec.3}, we present Metaverse architecture and its enabling technologies. In Section \ref{sec.4}, we discuss the similarities between Metaverse and blockchain development history, and the possibility that we end-up with multi fragmented Metaverses rather than a distributed universal Metaverse. Following that we present the proposed hybrid fog-edge architecture for Metaverse applications. In Section \ref{sec.5}, we present the simulation results of a use-case of distributed social Metaverse application that uses the proposed Fog-Edge architecture. Finally, we conclude the paper and outline future research directions in Section \ref{sec.6}

\section{Related work}
\label{sec.2}
Recently, we have witnessed an unprecedented research focus on Metaverse. In this section, we review the most recent researches about Metaverse and show the difference with our current work. 

Ning et al. \cite{ning2021survey} surveyed recent Metaverse development trend from five perspectives: network infrastructure, enabling technologies, services management, VR objects virtualization and virtual VR convergence. Wang et al. \cite{wang2022survey} survey Metaverse literature with special focus on security and privacy threats facing Metaverse applications, and reviewed the state-of-the-art privacy preserving solution for Metaverse applications. Similarly, Huynh-The et al. \cite{huynh2022artificial} explored the role of AI in Metaverse development, furthermore they discussed the application of machine learning and deep learning algorithms in Metaverse applications. Whereas Cheng et al. \cite{cheng2022will} briefly reviewed virtual reality social platforms that can be viewed as early prototypes of social Metaverse. Ynag et al. \cite{ynag2022fusing} surveyed the convergence of Artificial Intelligence (AI) and Blockchain technologies for Metaverse applications. They discussed how AI can empower blockchain based Metaverse applications such as virtual currency trading. Likewise, Jeon et al. \cite{jeon2022blockchain} discussed how AI and blockchain can affect Metaverse development in post COVID-19 era. Similarly, Gadekallu et al. \cite{gadekallu2022blockchain} discussed the potential applications of blockchain in Metaverse, specifically, data storage and acquisition,data interoperability, and blockchain for data privacy preservation. Nguyen et al. \cite{nguyen2021metachain} introduced MetaChain, a blockchain-based architecture that deals with usage of smart contracts operating on blockchain to manage and automate complex interactions between the Metaverse users and virtual service providers. Xu et al. \cite{xu2022full} discussed the potentials of AI, Edge computing and blockchain for ubiquitous, seamless access to the Metaverse. In the same vein, Lim et al. \cite{lim2022realizing} present the infrastructural architecture required for Metaverse with special focus on the convergence of edge intelligence and the infrastructure layer of the Metaverse. However, none of the above-mentioned works showed the potential of Fog and Edge computing for Metaverse application through practical use case as we will show in the rest of the paper.

\section{Metaverse architecture, technologies and applications}
\label{sec.3}
In this section, we discuss the main components of Metaverse applications, and its enabling technologies.

\begin{figure*}[!htbp]
	\centering
	\includegraphics[width=\textwidth]{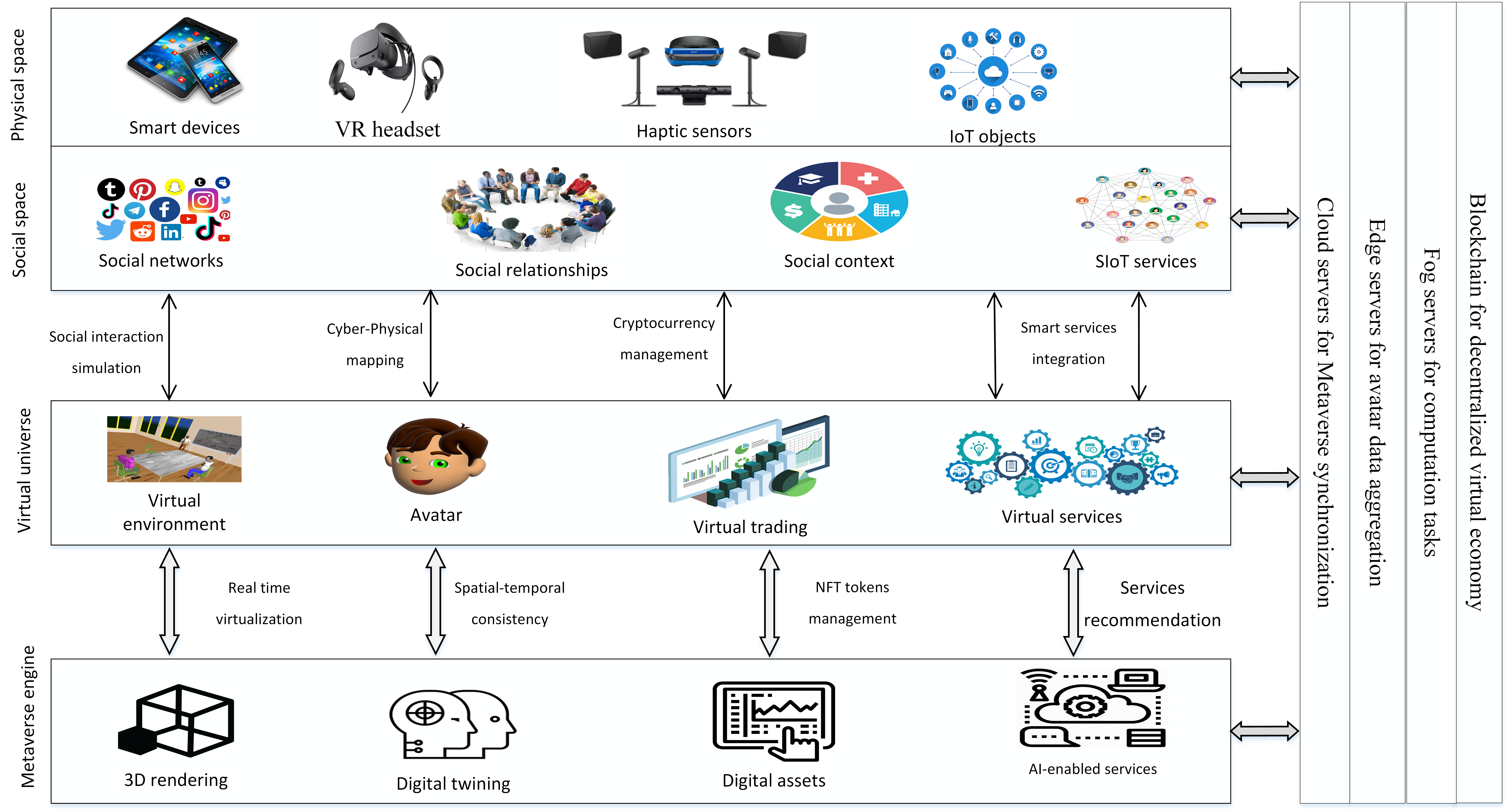}
	\caption{Physical-Social-Cyber mapping supported by Fog-Edge Computing and Blockchain decentralized virtual economy}
	\label{metaver_mapping}
\end{figure*}

\subsection{Architecture}

The Metaverse is the future big revolution of the Internet, and can be viewed as 3D version of the current Internet. As the users navigate and teleport in virtual worlds represented by their avatars, this requires the seamless mapping of users in the physical world and their counterparts in the virtual universe. The concept of physical-cyber mapping is not unique feature of Metaverse, it has been used in many other technologies, such as Cyber-Physical-Systems \cite{Dhelim2020entity}, Digital Twin \cite{duan2021metaverse}, Human-in-the-loop \cite{walsh2018human}, Hybrid Human-Artificial Intelligence \cite{wang2021survey} and human-robot interaction \cite{CAI2021505} to name a few, which will be inseparable with Metaverse. However, the Metaverse extends that to larger scale by mapping and simulating all our daily life activities to the cyberspace, and enriching such mapping with an immerse and interactive user experience.  To achieve this purpose, the Metaverse operating in multi-layer architecture as depicted in Figure \ref{metaver_mapping}.

\textbf{Physical space}: The access point to the Metaverse, this includes any device that enable the user to interact with the Metaverse. General-purpose mobile devices and computer can serve as Metaverse access points, but when accessing from such devices the users cannot benefit from many services and features offered by Metaverse applications. An immerse and interactive user experience can be offered only through XR devices and haptic sensors, which include sensors and headsets of virtual reality, augmented reality, mixed reality and extended reality. Metaverse applications are connected to IoT smart services, and actions that took place in Metaverse may trigger IoT smart services in the physical space \cite{dhelim2016stlf}. For instance, the temperature in current position of Metaverse avatar can be mapped with the temperature of IoT-enabled smart home air conditioning system to simulate the Metaverse weather \cite{dhelim2018cyber}.

\textbf{Social space}: Among the criticism to Metaverse, some may claim that the Metaverse is just a 3D social network enhanced by VR technologies. While such claim is valid up to some extent, in the sense that the Metaverse will transformed the way we interact with social networks, from single access point interaction to 3D world navigation accessed by VR technologies. However, Metaverse incorporate many concepts that go beyond traditional social networks, such as virtual assets trading, virtual learning environment, virtual military training, to name a few. In addition to the conventional social networks, the Metaverse is integrated with users' social context provided by Social Internet of Things (SIoT) services \cite{khelloufi2020social}. SIoT services are customized according to the user's social features; for example personality-aware recommendation systems \cite{Dhelim2021survery} and user interest mining based services \cite{Dhelim2020compath,Dhelim2020product} can offer social enhanced SIoT services.

\textbf{Virtual universe}: The physical and social space entities are mapped into virtual space. The users are represented by avatars, while other physical entities are represented by digital assets. The map of virtual world might be a representation of real world map, such as the world map or a specific building map, or an imaginary map like the maps used in multiplayer online games. Cyber-physical and digital twin application are integrated with Metaverse application to offer realistic cyber representation of the physical world.

\textbf{Metaverse engine}: It performs computations required to run virtual universe simulation, which includes the computational of heavy tasks such as collision detection in virtual universe and computation of 3D physics, and also other aspect of virtual universe that demand high computational power. For example, the validation of Non Fungible Token (NTF) trading transaction requires tremendous computational power, not to mention data processing of digital twin applications, and AI-enabled services like storytelling and recommendation services that empowers the virtual universe simulation. Current state-of-the-art Metaverse implementations perform the computational on the cloud, which limits the simulation capacity, and increase access latency.

\subsection{Enabling technologies}
Metaverse is the next version of 3D Internet that realizes a seamless integration of immersive and interoperable ecosystems navigable by user-controlled avatars. That requires large-scale mapping of entities from physical and social spaces to virtual universe. To do that, Metaverse relies on plethora of services backed by various cutting-edge technologies. In the following, we list out Metaverse enabling technologies.\\

\textbf{Blockchain} The Metaverse economy involve trading virtual assets and in some cases that involves interaction with real economy. Securing and managing the Big data resulted from virtual goods and virtual services trading requires persistent and transparent trading mechanisms. The blockchain technology can play an essential role in this regard. Blockchain can be leveraged to secure virtual trading and preserve the value of digital assets and services. Blockchain as a distributed solution can mitigate the drawback of centralized trading paradigm. One of the most promising applications of blockchain in Metaverse is NFT trading. NFT is used to mark the ownership of digital goods and assets. Utilizing blockchain in NFT trading can facilitate peer-to-peer digital assets exchange in a distributed and decentralized manner.

\textbf{AI} technologies empowers Metaverse in many aspects. Deep learning and machine learning algorithms have been proven effective tackling complicated problems. These algorithms perform the best when they learn from large amount of data, and in this regards, Metaverse activities can generate very large data. AI plays an essential role in creating realistic virtual 3D worlds and automate the content generation of the Metaverse. For instance, NVIDIA introduced GANverse3D, an AI-based framework that allow content creators to take photos of real objects and automatically generate virtual replicas with lights, physics models and PBR materials. Furthermore, deep learning libraries like Facebook's PyTorch3D or NVIDIA's TensorRT can be used for 3D objects rendering, which accelerate virtual scenes creation and minimize the computational cost. Facebook recently created AI research supercluster (RSC) \cite{Facebook2022}, which considered among the largest AI-enable supercomputer dedicated to the applications of AI for creating Metaverse scenes and content.

\textbf{B5G/6G} Navigating the Metaverse smoothly through VR technology depends mainly on the network latency. As VR technologies are delay-sensitive and required very short latency, communicates with Metaverse servers plays a pivotal role in this regard. Metaverse activities generate massive big data such as social communications between millions of users, the transmission of high resolution interactive 3D animations, which requires high network bandwidth. Therefore, beyond 5G (B5G) and 6G networks might partially fulfil Metaverse communication requirements. B5G and 6G enable a real-time, ubiquitous, and ultra-reliable communications for massive Metaverse devices with support for device-mobility, which can reach 1020 Gbps. \cite{bujari2020addressing}.

\textbf{XR technologies} XR refer to virtual visualization technologies, that includes virtual reality, mixed reality and augmented reality. XR head-mounted helmets and displays are the main access point to Metaverse. XR devices allow users to enjoy real-time immerse interactive experience through multi-sensory large-scale 3D modelling \cite{xi2022challenges}. XR devices and haptic sensors create a visual realistic virtual environment, while IoT smart sensors carry out environment digitization by sensing the surrounding objects. In this way, the users are not limited to screen-based devices like laptops or smartphones to access the Metaverse.

\section{A Proposal for a distributed Metaverse}
\label{sec.4}

\subsection{Universal Metaverse or interconnected Metaverses}

Most of the recent propositions of Metaverse design do not come in line with the vision of universal Metaverse. It is rather several independent and fragmented Metaverses that rely on different hardware and software technologies. The isolated implementations of independent Metaverses jeopardises its integration, and makes it extremely difficult to interconnect these independent Metaverses to form a universal Metaverse. This scenario is analogous to the history of Blockchain development. The early vision was a universal Blockchain, where the authority is distributed among public nodes and anyone can participate in decision-making process.  The universal Blockchain was proposed to solve the drawbacks of centralized authority model. However, since the launch of first cryptocurrency Bitcoin blockchain, we have witnessed a proliferation of thousands of independent cryptocurrency blockchains. Some of which offer the ability to host other cryptocurrencies or other distrusted applications to operate in their blockchain platform. Such as Ethereum open-source and decentralized blockchain \cite{buterin2014next}, which allows third parties to host an immutable and decentralized applications into its blockchain, these applications are governed by smart-contract. This blockchain service model is known as Blockchain-As-A-Service (BAAS) \cite{weerasinghe2021novel}; however, it is far from being a universal Blockchain, as it is still always dependent on the blockchain provider.  Moreover, many organizations have their own controllable blockchains; this kind is known as private blockchain due to the fact that it is controlled by a single authority or/and being accessible but limited entities. As one can notice, the development trend of blockchain has diverted from the original vision of a unified universal Blockchain. Even though in the last few years, there were several attempts to interconnect all the public blockchains into a single blockchain, but up to nowadays there is no prominent technology that allow such integration. That is due to the homogeneous algorithms and technologies that manages these independent blockchains. The current development of Metaverse is taking a similar track. As many companies have their own centralized Metaverse that is accessible only by their users. However, unlike blockchain, in these private Metaverses the users do not participate in the decision-making process. And it is barely a virtual social world that replace traditional social networks, rather than a universal Metaverse that replace the physical world.

\subsection{Distributed Metaverse}

\begin{figure}[!htbp]
	\centering
	\includegraphics[width=\columnwidth]{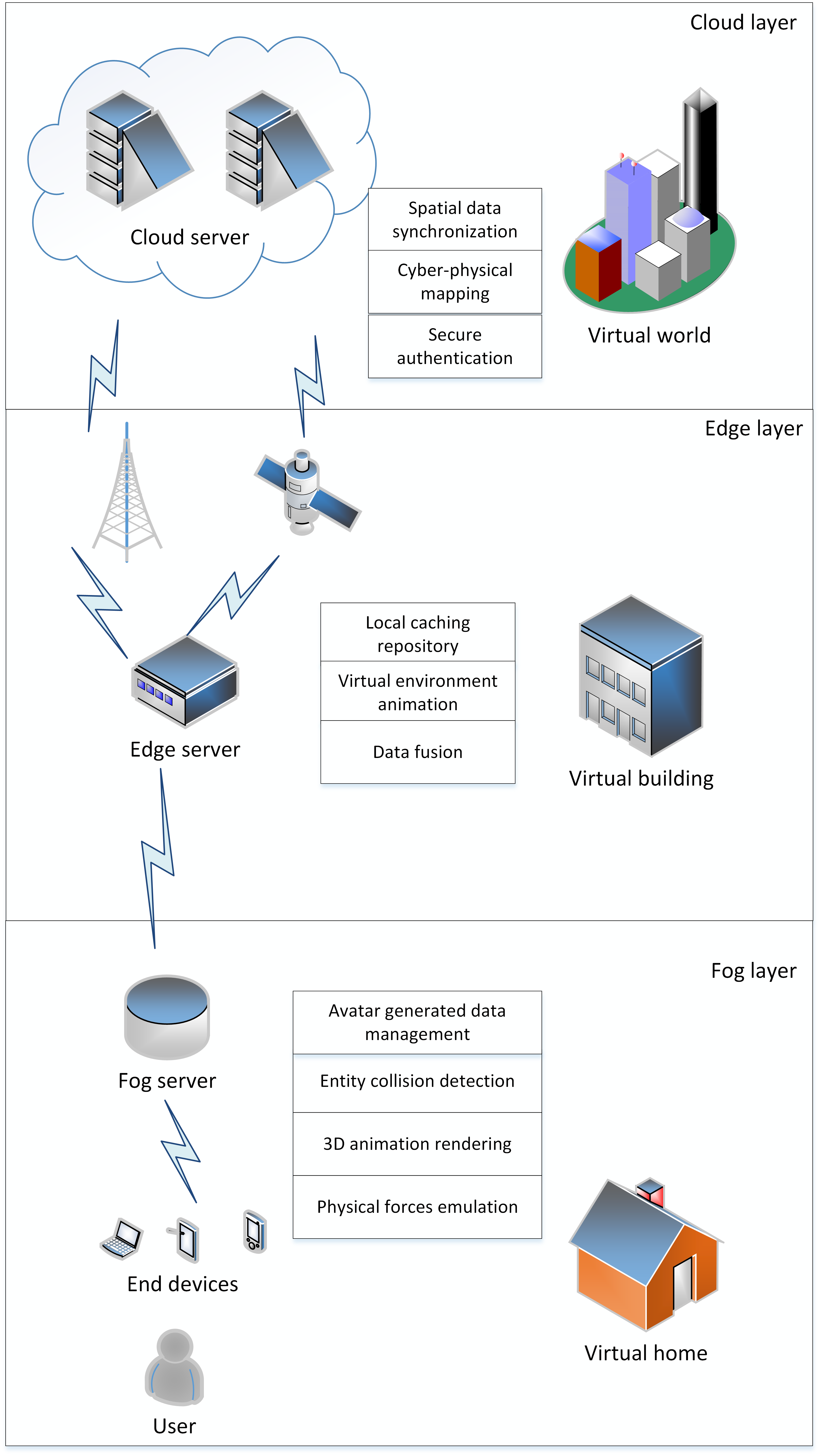}
	\caption{The proposed hybrid Fog-Edge architecture for Metaverse applications}
	\label{fog}
\end{figure}

The main driving factor behind blockchain fragmentation mentioned-above is the seek of authority and the benefits that such authority brings along. For instance, the inventor of a Bitcoin Satoshi Nakamoto possess around 1 million bitcoins, with a net worth of 73 billion US dollars, making it the 15${}^{th}$ richest person worldwide. Similarly, although a cloud-based Metaverse is unfavourable in term of visualization quality and also does not serve the vision of a universal Metaverse. However, such design gives organizations that have launched their own Metaverse the upper hand by controlling users' Metaverse data in their servers, and we will end up with thousands of separated Metaverses rather than a universal Metaverse. For example, Second Life's has many virtual worlds that are using centralized architecture, where a virtual world is divided into smaller regions, each managed by dedicated servers. Most of the computational required for running the virtual world simulations such as physics emulating, 3D animation and collision detection are performed by the region's centralized server. Therefore, the number of users that can access each region is limited by the server's computational and communication capacity. Such limitation defeats the purpose of virtual world, which as supposed to accommodate avatar as much as the real world location can physically accommodate people.  The Metaverse fragmentation problem and computational bottleneck can be resolved by considering a universal Metaverse where the data control and computation requirement of each entity in the virtual world is the responsibility of its corresponding entity in the physical space. In this context, we propose a distributed architecture that can achieve a universal Metaverse, and solves the computational bottleneck.  The generic data of the Metaverse such as virtual world spatial data and environment context information are publicly accessible by any entity. While any personal data about virtual entities are managed by its counterpart in the physical world. Furthermore, the computational cost of each entity such as collision detection or physics emulation are performed at the end-device of that physical entity. For example, for each avatar in the Metaverse, the user's edge devices are responsible to perform computation tasks related to the avatar's physical movement such as its momentum, mass and physical forces of surrounding entities. Such distributed architecture requires reliable independent infrastructure that manage the virtual world and synchronize and update events to all related entities' end devices. In Figure \ref{fog}, cloud servers are dedicated for virtual universe simulation, while the Edge servers are leveraged to perform computational task of specific region of the virtual world grid. Such regions could represent medium-level institution, for example, a Metaverse application that simulate virtual workplaces, the company have managed the employees' data. Finally, Fog servers perform computational task near end users in a smart home environment for instance. The advantage of layered architecture is twofold. Firstly, they users control their data, which enables organizations to access a universal Metaverse, rather than multiple separated Metaverses. Secondly, the computational bottleneck is resolved by distributing the computational cost of heavy tasks. 

\section{Use Case: Distributed Social Metaverse}
\label{sec.5}

\begin{figure}[!htbp]
	\centering
	\includegraphics[width=\columnwidth]{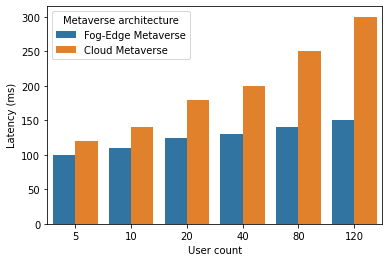}
	\caption{Metaverse access latency with various user count}
	\label{result1}
\end{figure}

\begin{figure}[!htbp]
	\centering
	\includegraphics[width=\columnwidth]{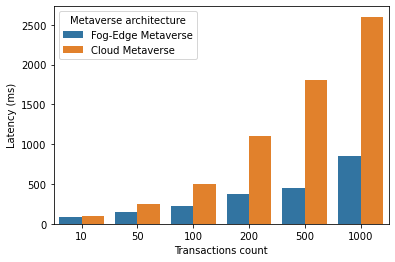}
	\caption{Metaverse access latency with various transaction count}
	\label{result2}
\end{figure}

To show the effectiveness of a distributed Metaverse architecture described in the last section, we have implemented a distributed social Metaverse application where the user data and computational tasks are performed at edge devices. In the proposed distributed social Metaverse, the users navigate a virtual map, and communicate with nearby users through text messaging; additional, they can place orders to buy digital assets. To simulated the network architecture, we have used iFogSim simulator \cite{gupta2017ifogsim}. iFogSim offer devices and fog/edge/cloud servers simulation. However, it does not support application specific scenario such as social network simulation, or blockchain management. We have extended iFogSim to simulate our Metaverse application. Specifically, we have created User class to simulate users and their properties through a contextual spatial information in the Metaverse. Blockchain/Block/Transaction classes are used for digital assets trading in Metaverse, and few other classes for events management. Every user's personal information such as messages and Metaverse navigation history are stored in its end-devices. While the digital asset trading history is stored in public blockchain. To simulate the required computational cost for 3D simulation, each user is associated with computational task. In our proposed architecture, these tasks are performed at the local fog server for spatial navigation, and at the edge server for social interaction management. We set cloud-based architecture as baseline, where all the computation and data management is performed at the cloud server. Figure \ref{result1} shows the latency difference between the two schemes with various Metaverse users count. As we can clearly observe, the latency significantly increases in cloud-based Metaverse with the increase of user count. Unlike Fog-Edge Metaverse, which maintain shortest latency, even when the user count increases. That is because in the latter, spatial navigation and collusion detection tasks are performed near the end-user, where in the former these tasks are performed on the cloud server, which increases the latency. Similarly, in Figure \ref{result2}, we see that latency exponentially increases in cloud-based Metaverse along with the increase of digital assets purchases transactions count. While Fog-Edge-Metaverse maintains a relatively low latency even with high transaction count, that is because the transactions are validated in edge sever.

\section{Discussion and open issues}
\label{sec.6}
With the expected widespread of Metaverse applications, there are many challenges that face Metaverse development:
\subsection{User safety}
Conventional social media companies still struggling to provide safe online environment and protect their users from online threats, including sexual harassment, cyber-bullying and identity theft. Metaverse is no exception to online threats, furthermore, the interactiveness of 3D virtual world had aggravated the situation, as the perpetrators can now access more sensitive information that they cannot get on traditional social networks. One of the most serious threat in Metaverse is sexual harassment, which had always existed in social networks, but Metaverse adds an extra layer of realistic visualization that worsen the situation. Metaverse users that experience sexual harassment in the virtual universe can suffer from psychological effects as a result of such virtual harassment. For example, the Belgian police have investigated a case of woman that has been virtually raped in Second Life \cite{Duranske2007}. Another case of a 7 years old girl been gang raped by two male avatars in Roblox \cite{Lanier2018}. Some may claim that what happened in the virtual universe is not real after all. However, recent studies showed that cyber sexual violence is strongly associated with symptoms of anxiety, stress, depression and post-traumatic reactions \cite{cripps2018cyber}. Metaverse companies have already imposed restrictions to mitigate sexual harassment. Setting restrictive measures can help preventing virtual sexual harassment incidents, however, it would defeat the purpose of social Metaverse where the avatar can interact without restrictions. For example, Meta have introduced personal boundary feature \cite{Robertson2022}, if activated, other avatars cannot approach your avatar less than two feet distance. Sexual harassment never been limited to physical assaults in the real world, and it will neither be in Metaverse. An AI-based solution is required to tackle this problem, without limiting social contact, yet being capable to detect and stop sexual harassment. In this regard, social artificial intelligence (ASI) \cite{asi} can play an important role, as it is more focused on users' social behaviors rather than their avatar movement patterns.

\subsection{Information privacy}

The Metaverse incorporates rich data about users more than traditional social media. The user's virtual identity is mapping of its real identity in the virtual world \cite{falchuk2018}, therefore protecting the information of an avatar in the virtual universe is as important as in physical world. Since the Metaverse leverages many technologies, the weakness in any technology threaten the whole system. For example, public blockchain transactions of digital assets purchases can be analysed to guess the identity of the buyers and sellers, which gives a golden chance for privacy intruders and hackers to target profitable victims.  Any leaking of personal information of avatars can leads to identity theft. Implementing multi-layer identity management for Metaverse users is paramount requirement to ensure avatars security and privacy.

\subsection{Metaverse addiction}

The excessive use of technology, specifically online gaming and internet can lead to an addition disorder known as Cyber-syndrome \cite{Ning2018}. The Metaverse is no exception of such addition, furthermore recent studies showed that VX technologies can increase the risk of internet addition by 44\% compare to traditional access devices such as smartphone or laptop {\textbackslash}cite$\mathrm{\{}$rajan2018virtual$\mathrm{\}}$.  As many activities will take place in the Metaverse, ranging from virtual education and virtual workplace to virtual parties. Spending long hours hearing XR helmets and glasses that are few inches from users' eyes will cause severe eye problems. Users may find difficulties to separate reality from virtual universe, let alone the impact of light-intensive images on our eyes. A recent study showed that kids who used VR glasses for 20 minutes had difficulties in distinguishing the distances of objects in reality as a result of vergence-accommodation conflict. \cite{yamada2017children}.

\subsection{Evaluation of user experience}

In the proposed solution we envision the execution of some Metaverse components in the fog/edge servers and in the end-device instead of in the central cloud. These are mostly related to user avatar data management, collision detection, rendering and management of the interaction with the physical world. This approach requires the transmission of real-time data from other Metaverse physical entities (objects and users) for the relevant processing. This is the case for instance in case of interaction among avatars of people located in quite far ends of the Internet. Where to process and store the data and which part of the required processing  to execute and at which temporal and spatial accuracy have to be decided by taking into account the final quality delivered to the users. This call for appropriate experience models that estimate the delivered quality on the basis of introduces latency and rendering quality influencing factors. These models still have to be studied as only preliminary solutions have been defined for VR/AR communications \cite{9658887}.

\section{Conclusion}
\label{sec.8}
The Metaverse will replace the traditional Internet and dominate all aspects of our lives. The state-of-the-art cloud-based Metaverse architecture can limit virtualization latency. In this paper, we have proposed a hybrid Fog-Edge computing architecture for Metaverse applications, the proposed architecture make use of edge devices to fulfil the required computational for heavy tasks such as collision detection in virtual universe and computation of 3D physics. the computational tasks related to a virtual entity are performed at the end-device of that physical entity. Simulation results showed that the proposed architecture can reduce visualization latency by 50\% compared to the legacy cloud-based Metaverse applications.

\section*{Acknowledgment}
This work was supported by the National Natural Science Foundation of China under Grant 61872038.

\ifCLASSOPTIONcaptionsoff
  \newpage
\fi

\bibliographystyle{IEEEtran}

\bibliography{refs}

%

\begin{IEEEbiography}[{\includegraphics[width=1in,height=1.25in,clip,keepaspectratio]{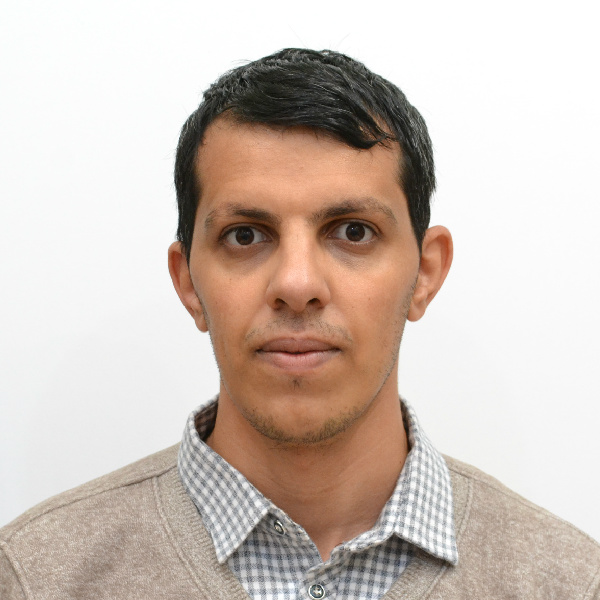}}]{Sahraoui Dhelim}
	is a postdoctoral researcher in University College Dublin, Ireland. He was a visiting researcher in Ulster University, UK (2020-2021). He obtained his PhD degree in Computer Science and Technology from University of Science and Technology Beijing, China, in 2020. And a Masters degree in Networking and Distributed Systems from the University of Laghouat, Algeria, in 2014. He serves as workshop chair of Cyberspace congress (CyberCon). His research interests include Social Computing, User Modeling, Deep-learning, Recommendation Systems and Intelligent Transportation Systems.
\end{IEEEbiography}

\vskip 0pt plus -1fil
\begin{IEEEbiography}[{\includegraphics[width=1in,height=1.25in,clip,keepaspectratio]{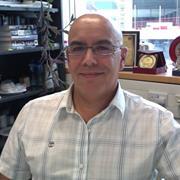}}]{Mohand Tahar Kechadi}
is a full professor in school of computer science, University College Dublin, Ireland. He received master’s and Ph.D. degrees in computer science from the University of Lille 1, France. His research interests include data mining, distributed data mining heterogeneous distributed systems, grid and cloud computing, and digital forensics and cyber-crime investigations. He is a member of the Communications of the ACM journal and IEEE Computer Society. He is an Editorial Board Member of journal of Future Generation Computer Systems.
\end{IEEEbiography}

\vskip 0pt plus -1fil

\begin{IEEEbiography}[{\includegraphics[width=1in,height=1.25in,clip,keepaspectratio]{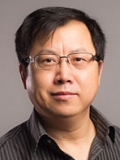}}]{Liming Chen}
	is a professor in the School of Computer  Science  and  Informatics  at  University  of  Ulster,  Newtownabbey,  United  Kingdom.  He  received his  B.Eng  and  M.Eng  from  Beijing  Institute  of Technology  (BIT),  Beijing,  China,  and  his  Ph.D  in Artificial Intelligence from De Montfort University,UK.  His  research  interests  include  data  analysis,ubiquitous computing, and human-computer interaction. Liming is a Fellow of IET, a Senior Member of IEEE, a Member of the IEEE Computational Intelligence Society (IEEE CIS), a Member of the IEEE CIS Smart World Technical Committee (SWTC), and the Founding Chair of the IEEE CIS SWTC Task Force on User-centred Smart Systems (TF-UCSS). He has served as an expert assessor, panel member and evaluator for UK EPSRC (Engineering and Physical Sciences Research Council, member of the Peer Review College), ESRC (Economic and Social Science Research Council), European Commission Horizon 2020 Research Program, Danish Agency for Science and Higher Education, Denmark, Canada Foundation for Innovation (CFI), Canada, Chilean National Science and Technology Commission (CONICYT), Chile, and NWO (The Netherlands Organisation for Scientific Research), Netherlands.
\end{IEEEbiography}

\vskip 0pt plus -1fil
\begin{IEEEbiography}[{\includegraphics[width=1in,height=1.25in,clip,keepaspectratio]{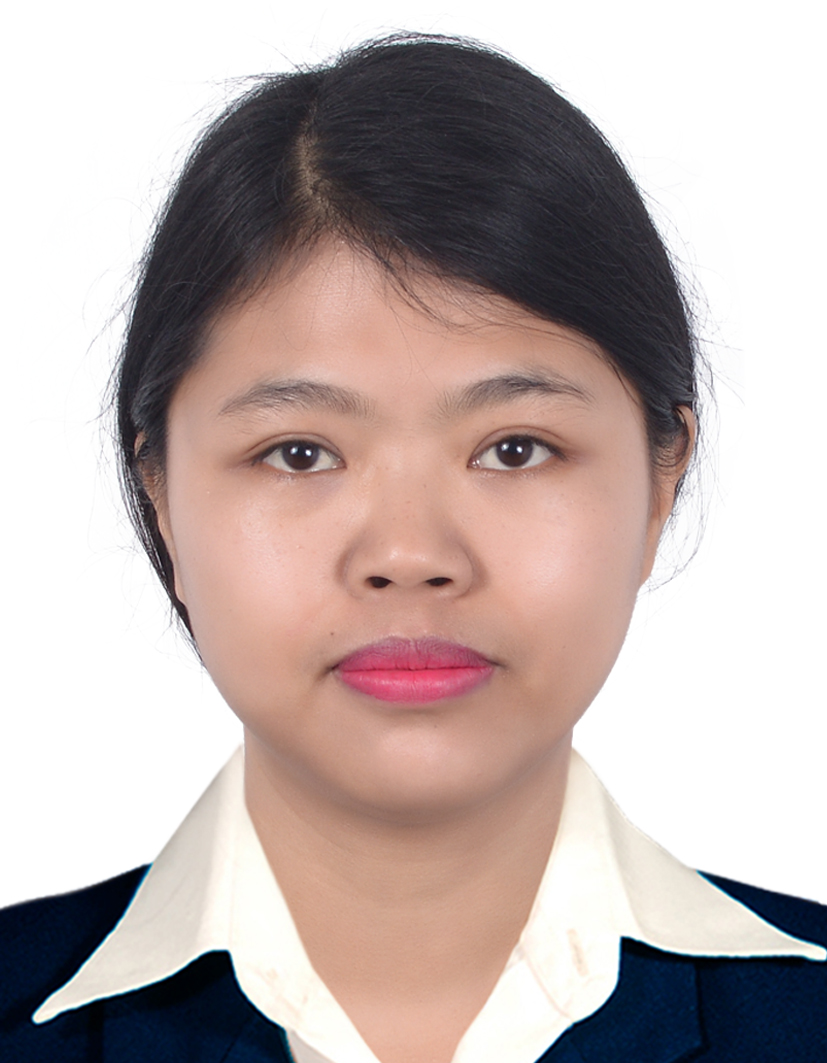}}]{Nyothiri Aung}
	is a postdoctoral researcher in University College Dublin, Ireland.
	She received her PhD in Computer Science and Technology from University of Science and Technology Beijing, China, 2020.
	 and a Master of Information Technology form Mandalay Technological University, Myanmar, 2012. She worked as a tutors at the Department of Information Technology in Technological University of Meiktila, Myanmar (2008-2010). And System Analyst of ACE Data System, Myanmar (2012-2015). Her research interests include Social Computing, Personality Computing and Intelligent Transportation System.
\end{IEEEbiography}	

\vskip 0pt plus -1fil

\begin{IEEEbiography}[{\includegraphics[width=1in,height=1.25in,clip,keepaspectratio]{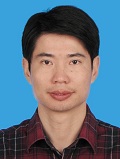}}]{Huansheng Ning}
	Received his B.S. degree from Anhui University in 1996 and his Ph.D. degree from Beihang University in 2001. Now, he is a professor and vice dean of the School of Computer and Communication Engineering, University of Science and Technology Beijing, China. His current research focuses on the Internet of Things and general cyberspace.
	He is the founder and chair of the Cyberspace and Cybermatics International Science and Technology Cooperation Base.
	He has presided many research projects including Natural Science Foundation of China, National High Technology Research and Development Program of China (863 Project). He has published more than 150 journal/conference papers, and authored 5 books. He serves as an associate editor of IEEE Systems Journal (2013-Now), IEEE Internet of Things Journal (2014-2018), and as steering committee member of IEEE Internet of Things Journal (2016-Now).
\end{IEEEbiography}

\vskip 0pt plus -1fil

\begin{IEEEbiography}[{\includegraphics[width=1in,height=1.25in,clip,keepaspectratio]{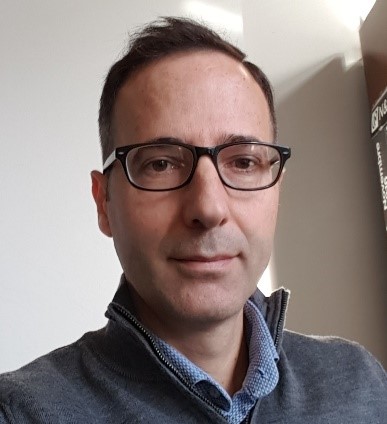}}]{Luigi Atzori} is Full Professor at the Department of Electrical and Electronic Engineering, University of Cagliari (Italy) and Research Associate at the Multimedia Communications Laboratory of CNIT (Consorzio Nazionale Inter-universitario per le Telecomunicazioni). Prof. Atzori received his Ph.D. degree in electronic engineering and computer science from the University of Cagliari in 2000. He spent the seven months from November 2003 to May 2004 at the Department of Electrical and Computer Engineering, University of Arizona, as Fulbright Visiting Scholar. L. Atzori research interests are in multimedia communications and computer networking and services in the Internet of Things and Social Internet of Things. L. Atzori is senior member of IEEE (since 2009) and has been the Steering Committee Chair of the IEEE Multimedia Communications Committee (MMTC) for the years 2014-2016. He has been the associate and guest editor for several journals, included: ACM/Springer Wireless Networks Journal, IEEE IoT journal, IEEE Comm. Magazine, the Springer Monet Journal, Elsevier Ad Hoc Networks, and the Elsevier Signal Processing: Image Communications Journal. Currently he serves in the editorial board of the following journals: Elsevier Digital Communications and Networks and IEEE Open Journal of the Communications Society. He served as a technical program chair for various international conferences and workshops, including ICC and Globecom workshops, ACM MobiMedia and VLBV. He served as a reviewer and panelist for many funding agencies, including FP7, Horizon2020, Cost Actions, Italian MIUR and Regional funding agency.
\end{IEEEbiography}
\end{document}